\documentclass[aps,twocolumn,prd,groupedaddress,nofootinbib,amssymb,eqsecnum,epsfig]{revtex4}
\usepackage{graphicx}
\usepackage{bm}
\usepackage{amsmath}
\usepackage{color}
\usepackage{amsfonts}

\begin{document}
\newcommand{\newc}{\newcommand}

\newcommand{\ben}{\begin{eqnarray}}
\newcommand{\een}{\end{eqnarray}}
\newc{\be}{\begin{equation}}
\newc{\ee}{\end{equation}}
\newc{\ba}{\begin{eqnarray}}
\newc{\ea}{\end{eqnarray}}
\newc{\bea}{\begin{eqnarray*}}
\newc{\eea}{\end{eqnarray*}}
\newc{\D}{\partial}
\newc{\ie}{{\it i.e.} }
\newc{\eg}{{\it e.g.} }
\newc{\etc}{{\it etc.} }
\newc{\etal}{{\it et al.}}
\newcommand{\nn}{\nonumber}
\newc{\ra}{\rightarrow}
\newc{\lra}{\leftrightarrow}
\newc{\lsim}{\buildrel{<}\over{\sim}}
\newc{\gsim}{\buildrel{>}\over{\sim}}

\title{Beyond generalized Proca theories}

\author{
Lavinia Heisenberg$^{1}$, 
Ryotaro Kase$^{2}$, and
Shinji Tsujikawa$^{2}$ }

\affiliation{
$^1$Institute for Theoretical Studies, ETH Zurich, Clausiusstrasse 47, 8092 Zurich, Switzerland\\
$^2$Department of Physics, Faculty of Science, Tokyo University of Science, 1-3, Kagurazaka,
Shinjuku-ku, Tokyo 162-8601, Japan}

\date{\today}

\begin{abstract}

We consider higher-order derivative interactions beyond second-order 
generalized Proca theories that propagate only the three 
desired polarizations of a massive vector field besides the two tensor 
polarizations from gravity. 
These new interactions follow the similar construction criteria to 
those arising in the extension of scalar-tensor Horndeski theories 
to Gleyzes-Langlois-Piazza-Vernizzi (GLPV) theories. 
On the isotropic cosmological background, we show the existence 
of a constraint with a vanishing Hamiltonian that removes 
the would-be Ostrogradski ghost.
We study the behavior of linear perturbations 
on top of the isotropic cosmological background in the presence of 
a matter perfect fluid and find the same number of propagating degrees of freedom 
as in generalized Proca theories (two tensor polarizations, two transverse 
vector modes, and two scalar modes). Moreover, we obtain the conditions for the avoidance 
of ghosts and Laplacian instabilities of tensor, vector, and scalar perturbations.  
We observe key differences in the scalar sound speed, which is mixed with 
the matter sound speed outside the domain of generalized Proca theories.

\end{abstract}


\maketitle

\section{Introduction}

General Relativity (GR) is still the fundamental theory
for describing the gravitational interactions even after a century. 
Cosmological observations \cite{SNIa,CMB,BAO} led to the 
standard model yielding an accelerated expansion of the late Universe 
driven by the cosmological
constant. The standard model of particle physics describes the strong
and electro-weak interactions with an exquisite experimental success 
marking the milestone in high-energy physics. It is still a big challenge
to unify gravity with the known forces in Nature and to merge these two
standard models into a single theory. Moreover, employing the usual
techniques of quantum field theory, we are not able to explain the small
observed value of the cosmological constant. 
On the other hand, this has motivated to consider infra-red modifications of gravity which could
account for an appropriate screening of the cosmological constant. 
On a similar footing, one can also consider infra-red gravitational 
modifications to realize an effective negative pressure against 
gravity in form of dark energy \cite{review}.

The simplest and mostly studied large-distance modification of gravity is attributed to 
an additional scalar field beyond the standard model of particle physics, 
e.g., the DGP braneworld \cite{DGP}, Galileons \cite{Galileon}, 
and massive gravity \cite{mgravity}. 
The scalar field arising in such theories can have
non-trivial self-interactions but also it can be generally 
coupled to gravity \cite{Amendola,coGa}.
These interactions have to be constructed with great caution to guarantee the
absence of ghost-like Ostrogradski instability \cite{Ostro}, 
which otherwise would yield an
unbounded Hamiltonian from below. 

It is well known that matter fields have to be coupled to the Lovelock invariants 
or to the divergence-free tensors constructed from the Lovelock invariants. 
Hence they can for instance couple to the volume element $\sqrt{-g}$ and 
to the Ricci scalar $R$ which are the only two non-trivial Lovelock invariants, 
since the Gauss-Bonnet term is topological in four dimensions. 
Furthermore, they can couple to the divergence-free
metric $g_{\mu\nu}$, Einstein tensor $G_{\mu\nu}$, and the double dual Riemann 
tensor $L_{\mu \nu \alpha \beta}$. In flat space-time the ghost-free scalar interactions
with derivatives acting on them are known as the Galileon interactions \cite{Galileon}. If one
would naively promote the partial derivatives to covariant derivatives, this procedure
would yield the equations of motion higher than second order \cite{Ostro}. 
The appearance of higher-order derivative terms can be avoided by introducing 
non-minimal couplings to gravity through the Lovelock invariants or 
the divergence-free tensors. 

Horndeski theories \cite{Horndeski} constitute the most general scalar-tensor 
interactions with second-order equations of motion. 
In these theories there is only one scalar degree of freedom (DOF) besides two 
graviton polarizations without having the Ostrogradski instability \cite{Horn2}.
It is a natural question to ask whether abandoning the requirement of second-order 
equations of motion inevitably alters the propagating DOF. 
Allowing interactions beyond the Horndeski 
domain will introduce derivative interactions higher than second order. 
However, this does not necessarily mean that the number of propagating DOF increases. 
Exactly this spirit was followed in GLPV theories \cite{GLPV}, where they expressed the
Horndeski Lagrangian in terms of the 3+1 Arnowitt-Deser-Misner (ADM) decomposition
of space-time in the unitary gauge \cite{Gleyzes13} and did not impose the two conditions that 
Horndeski theories obey. The Hamiltonian analysis in the unitary gauge revealed 
that there is still only one scalar DOF \cite{Hami}. 
The cosmology and the spherically symmetric solutions in GLPV 
theories have been extensively studied in Refs.~\cite{Gergely,Kase14,cosGLPV,spheGLPV}. 
The ghost freedom beyond the unitary gauge and beyond a conformal and disformal
transformation is still an ongoing research investigation 
in the literature \cite{counting,Domenech,Fujita,Langlois,Koyama,Zuma}. 

Even if the large-distance modifications of gravity through a scalar field are simpler, considerations in form
of a vector field can yield interesting phenomenology for the cosmic expansion and growth
of large-scale structures. Furthermore, the presence of the vector field might explain the 
anomalies reported in CMB observations \cite{Ade}.
For a gauge-invariant vector field, the only new interaction
is via a coupling of the field strength tensor to the double dual Riemann tensor. 
Unfortunately, the existence of derivative self-interactions similar to those arising for covariant Galileons 
is forbidden for a massless, Lorentz-invariant vector field coupled to gravity \cite{Mukoh}. 

However, this negative result does not apply to massive vector fields, for which one can successfully 
construct derivative self-interactions due to the broken $U(1)$ symmetry. 
The idea was to construct interactions with only three propagating degrees of freedom, out of which
two would correspond to the transverse and one to the longitudinal mode of the vector field. 
This was systematically constructed in Ref.~\cite{Heisenberg} 
together with the Hessian and Hamiltonian analysis. The key point is the requirement that 
the longitudinal mode belongs to the class of Galileon/Horndeski theories. 
This constitutes the generalized Proca theories up to the quintic Lagrangian on curved
space-time with second-order equations of motion, which is guaranteed by the presence of
non-minimal couplings to the Lovelock invariants in the same spirit as in the 
scalar Horndeski theories \cite{Heisenberg,Tasinato,Allys,Jimenez16}. 

One can also construct the sixth-order derivative interactions, if one allows for trivial interaction
terms for the longitudinal mode \cite{Allys,Jimenez16}. Its generalization to curved space-time
contains the double dual Riemann tensor, which 
keeps the equations of motion up to second order \cite{Jimenez16}. 
In fact, this sixth-order Lagrangian accommodates similar vector-tensor theories 
constructed by Horndeski in 1976 \cite{Horndeski2}. We refer the reader 
to Refs.~\cite{Barrow,Jimenez13,TKK,Fleury,Hull,Li,scvector,DeFelice16,Geff} 
for related works. The second-order massive vector theories up to the 
sixth-order Lagrangian studied in Refs.~\cite{Jimenez16,DeFelice16,Geff} constitute
the generalized Proca theories. 

It is a natural follow-up question to ask whether or not the extension of 
generalized Proca theories is possible 
in such a way that there are still three propagating vector DOF even with 
derivatives higher than second order.
In the GLPV extension of Horndeski theories, the Lagrangians of 
two additional scalar derivative interactions can be expressed 
in terms of the anti-symmetric Levi-Civita tensor. 
Outside the domain of generalized Proca theories, one can also construct generalized
Lagrangians by using the Levi-Civita tensor.
It is then expected that, in beyond-generalized Proca theories, 
the longitudinal vector mode would have some correspondence with 
the scalar mode in GLPV theories, but there will be also new interactions 
corresponding to the purely intrinsic vector modes. 

In this Letter, we will propose candidates for new beyond-generalized Proca Lagrangians
in Sec.~\ref{theorysec} to study the possibility of the healthy 
extension of generalized Proca theories.
In Sec.~\ref{Hessiansec} we derive the background equations of motion 
on the flat Friedmann-Lema\^{i}tre-Robertson-Walker (FLRW) background 
and the associated Hamiltonian ${\cal H}$. 
We see that, even in the presence of these new interactions, there exists
a second class constraint (${\cal H}=0$) that removes the Ostrogradski ghost.
In Sec.~\ref{FLRWsec1} we consider linear cosmological perturbations on the 
flat FLRW background and show that the number of DOF in beyond-generalized 
Proca theories is not altered relative to that in 
generalized Proca theories. We also study what kinds of differences arise for the stability of 
perturbations by extending generalized Proca theories to beyond-generalized Proca theories. 
Sec.~\ref{consec} is devoted to conclusions and future outlook.

\section{Extension of generalized Proca theories to beyond-generalized Proca theories}
\label{theorysec}

The generalized Proca theories are characterized by second-order interactions with two transverse and 
one longitudinal polarizations of a vector field $A^{\mu}$ coupled to gravity.
Introducing the field tensor $F_{\mu \nu}=\nabla_{\mu}A_{\nu}-\nabla_{\nu}A_{\mu}$, 
where $\nabla_{\mu}$ is the covariant derivative operator, 
the four-dimensional action of generalized Proca theories is given by 
\be
S_{\rm gen.Proca}=\int d^4 x \sqrt{-g} \sum_{i=2}^{6} {\cal L}_i\,,
\label{actionGP}
\ee
where $g$ is the determinant of the metric tensor $g_{\mu \nu}$, and 
\ba
{\cal L}_2 &=& G_2(X,F,Y)\,,
\label{L2}\\
{\cal L}_3 &=& G_3 \nabla_{\mu}A^{\mu}\,,
\label{L3}\\
{\cal L}_4 &=& 
G_4R+
G_{4,X} \left[ (\nabla_{\mu} A^{\mu})^2
-\nabla_{\rho}A_{\sigma}
\nabla^{\sigma}A^{\rho} \right]\,,\label{L4} \\
{\cal L}_5 &=& 
G_{5} G_{\mu \nu} \nabla^{\mu} A^{\nu}
-\frac16 G_{5,X} [ (\nabla_{\mu} A^{\mu})^3 \nonumber \\
&&-3\nabla_{\mu} A^{\mu}
\nabla_{\rho}A_{\sigma} \nabla^{\sigma}A^{\rho} 
+2\nabla_{\rho}A_{\sigma} \nabla^{\gamma}
A^{\rho} \nabla^{\sigma}A_{\gamma}] \nonumber \\
& &-g_5 \tilde{F}^{\alpha \mu}
{\tilde{F^{\beta}}}_{\mu} \nabla_{\alpha} A_{\beta}\,,
\label{L5}\\
{\cal L}_6 &=& G_6 L^{\mu \nu \alpha \beta} 
\nabla_{\mu}A_{\nu} \nabla_{\alpha}A_{\beta} \nonumber\\
&&+\frac12 G_{6,X} \tilde{F}^{\alpha \beta} \tilde{F}^{\mu \nu} 
\nabla_{\alpha}A_{\mu} \nabla_{\beta}A_{\nu}\,.
\label{L6}
\ea
The function $G_2$ depends on the following 
three quantities
\be
X=-\frac{A_{\mu} A^{\mu}}2 \,,\quad
F=-\frac{F_{\mu \nu} F^{\mu \nu}}4 \,,\quad
Y= A^{\mu}A^{\nu} {F_{\mu}}^{\alpha} 
F_{\nu \alpha}\,, 
\label{Xdef}
\ee
while $G_{3,4,5,6}$ and $g_5$ are arbitrary 
functions of $X$ with the notation $G_{i,X} \equiv \partial G_{i}/\partial X$. 
The vector field is coupled to the Ricci scalar $R$ and 
the Einstein tensor $G_{\mu \nu}$
through the functions $G_4(X)$ and $G_5(X)$.
The $L^{\mu \nu \alpha \beta}$ and 
$\tilde{F}^{\mu \nu}$ are the double dual Riemann 
tensor and the dual strength tensor
defined, respectively, by
\be
L^{\mu \nu \alpha \beta}=\frac14 {\cal E}^{\mu \nu \rho \sigma} 
{\cal E}^{\alpha \beta \gamma \delta} R_{\rho \sigma \gamma \delta}\,,
\quad
\tilde{F}^{\mu \nu}=\frac12 {\cal E}^{\mu \nu \alpha \beta}
F_{\alpha \beta}\,,
\ee
where $R_{\rho \delta \gamma \delta}$ is the Riemann tensor and 
${\cal E}^{\mu \nu \rho \sigma}$ is the Levi-Civita tensor 
obeying the normalization ${\cal E}^{\mu \nu \rho \sigma}{\cal E}_{\mu \nu \rho \sigma}=-4!$.
We can potentially include the dependence of the quantity 
$F^{\mu \nu} \tilde{F}_{\mu \nu}$ in the function $G_2$ \cite{Heisenberg,Fleury}. 
If we impose the parity invariance, however, it does not contribute 
to the perturbations at linear order, so we do not take into account 
such dependence in $G_2$.

The action (\ref{actionGP}) was constructed to keep the equations of motion 
up to second order to avoid the appearance of an extra DOF besides 
two transverse and one longitudinal modes of the vector field \cite{Heisenberg}.  
Each Lagrangian density can be expressed in terms of 
the Levi-Civita tensor ${\cal E}_{\mu_1 \mu_2 \mu_3 \mu_4}$ and 
the first derivatives of $A^{\mu}$.
The anti-symmetric property of ${\cal E}_{\mu_1 \mu_2 \mu_3 \mu_4}$ 
allows us to eliminate the terms containing time derivatives of the temporal 
vector component $A^0$, such that the additional DOF does not propagate.
In this set up the derivatives of $A^{\mu}$ higher 
than first-order are not taken into account, as they 
give rise to the derivatives of a scalar field $\pi$ 
higher than second order in the Lagrangian 
by taking the limit $A^{\mu} \to \nabla^{\mu}\pi$.

The action (\ref{actionGP}) consists of three parts. 
The first part corresponds to the Lagrangian densities 
(with the index $i=0,1,2,3$):
\be
{\cal L}_{i+2}^{{\rm Ga}} = g_{i+2}\,
\hat{\delta}_{\alpha_1 \cdots \alpha_i
\gamma_{i+1} \cdots \gamma_{4}}^{\beta_1 \cdots \beta_i
\gamma_{i+1} \cdots \gamma_{4}}
\nabla_{\beta_1} A^{\alpha_1}
\cdots \nabla_{\beta_i} A^{\alpha_i}\,,
\label{LG}
\ee
where $g_{i+2}$ are functions of $X$ and we have introduced the operator 
$\hat{\delta}_{\alpha_1 \cdots \alpha_i
\gamma_{i+1} \cdots \gamma_{4}}^{\beta_1 \cdots \beta_i
\gamma_{i+1} \cdots \gamma_{4}}={\cal E}_{\alpha_1 \cdots \alpha_i
\gamma_{i+1} \cdots \gamma_{4}}  {\cal E}^{\beta_1 \cdots \beta_i
\gamma_{i+1} \cdots \gamma_{4}}$. They
recover those of Minkowski Galileons from the scalar part $\pi$ of 
$A^{\mu}$ for the functions $g_{2,3,4,5} \propto X$.
The second part arises from the terms derived by exchanging some 
of the indices in ${\cal L}_{4,5}^{\rm Ga}$, i.e., 
\ba
\hspace{-1.2cm}
&&{\cal L}_4^{V} = \tilde{h}_{4}\,
\hat{\delta}_{\alpha_1 \alpha_2\gamma_3\gamma_4}^{\beta_1 \beta_2\gamma_3\gamma_4}
\nabla_{\beta_1}A_{\beta_2}\nabla^{\alpha_1}A^{\alpha_2}\,,\\
\hspace{-1.2cm}
&&{\cal L}_5^{V} = \tilde{h}_{5}
\hat{\delta}_{\alpha_1 \alpha_2\alpha_3\gamma_4}^{\beta_1 \beta_2\beta_3\gamma_4}
\nabla^{\alpha_1} A^{\alpha_2}
\nabla_{\beta_1} A_{\beta_2} \nabla^{\alpha_3} A_{\beta_3}\,,
\ea
with again $\hat{\delta}_{\alpha_1 \alpha_2\gamma_3\gamma_4}^{\beta_1 \beta_2\gamma_3\gamma_4}={\cal E}_{\alpha_1 \alpha_2\gamma_3\gamma_4}
{\cal E}^{\beta_1 \beta_2\gamma_3\gamma_4}$ and general functions $\tilde{h}_{4}$ and $\tilde{h}_{5}$ depending on $X$. These interactions can be regarded as the intrinsic vector modes that vanish 
in the scalar limit $A^{\mu} \to \nabla^{\mu}\pi$. 
The Lagrangian density ${\cal L}_6$ contains the intrinsic vector 
contribution
\be
{\cal L}_6^{V}
=\tilde{h}_{6}\,
\hat{\delta}_{\alpha_1 \alpha_2 \alpha_3 \alpha_4}^{\beta_1 \beta_2\beta_3\beta_4}
\nabla_{\beta_1} A_{\beta_2} \nabla^{\alpha_1}A^{\alpha_2}
\nabla_{\beta_3} A^{\alpha_3} \nabla_{\beta_4} A^{\alpha_4}\,.
\label{L6V}
\ee
The third part corresponds to the non-minimal coupling terms 
$G_4(X)R$, $G_{5}(X) G_{\mu \nu}\nabla^{\mu}A^{\nu}$, 
and $G_6(X) L^{\mu \nu \alpha \beta} 
\nabla_{\mu}A_{\nu} \nabla_{\alpha}A_{\beta}$, which are 
required to keep the equations of motion up to second order \cite{Heisenberg,Jimenez16}.

If we try to make the minimal extension of the above generalized Proca theories, 
we can take into account terms containing the products of 
$A^{\alpha_1}A_{\beta_1}$ and the first derivatives of $A^{\mu}$. 
Let us consider the following new Lagrangian densities
\ba
\hspace{-1.2cm}
& &{\cal L}_4^{\rm N}
=f_4 \hat{\delta}_{\alpha_1 \alpha_2 \alpha_3 \gamma_4}^{\beta_1 \beta_2\beta_3\gamma_4}
A^{\alpha_1}A_{\beta_1}
\nabla^{\alpha_2}A_{\beta_2} 
\nabla^{\alpha_3}A_{\beta_3}\,, \label{L4N}\\
\hspace{-1.2cm}
& &{\cal L}_5^{\rm N}
=
f_5 \hat{\delta}_{\alpha_1 \alpha_2 \alpha_3 \alpha_4}^{\beta_1 \beta_2\beta_3\beta_4}
A^{\alpha_1}A_{\beta_1} \nabla^{\alpha_2} 
A_{\beta_2} \nabla^{\alpha_3} A_{\beta_3}
\nabla^{\alpha_4} A_{\beta_4}\,,\label{L5N} \\
\hspace{-1.2cm}
& &\tilde{{\cal L}}_5^{\rm N}
=
\tilde{f}_{5}
\hat{\delta}_{\alpha_1 \alpha_2 \alpha_3 \alpha_4}^{\beta_1 \beta_2\beta_3\beta_4}
A^{\alpha_1}A_{\beta_1} \nabla^{\alpha_2} 
A^{\alpha_3} \nabla_{\beta_2} A_{\beta_3}
\nabla^{\alpha_4} A_{\beta_4}\,,
\label{L5Nd}
\ea
with the functions $f_{4,5}$ and $\tilde{f}_{5}$ depending on $X$.
If we take the limit $A^{\mu} \to \nabla^{\mu} \pi$, 
the Lagrangian densities ${\cal L}_4^{\rm N}$ and 
${\cal L}_5^{\rm N}$ for the scalar field $\pi$ 
are equivalent to those appearing in GLPV theories \cite{GLPV}.
Thus, the above construction of new derivative interactions 
is analogous to the GLPV extension of scalar Horndeski theories, 
but in our case the situation is more involved due to the existence 
of transverse vector modes. 
We also need to take into account the intrinsic vector term 
$\tilde{{\cal L}}_5^{\rm N}$ derived after exchanging 
the indices $\beta_2$ and $\alpha_3$ in ${\cal L}_5^{\rm N}$.
Note, that we did not include the term $\tilde{{\cal L}}_4^{\rm N}
=\tilde{f}_4 \hat{\delta}_{\alpha_1 \alpha_2 \alpha_3 \gamma_4}^{\beta_1 \beta_2\beta_3\gamma_4}
A^{\alpha_1}A_{\beta_1}\nabla^{\alpha_2}A^{\alpha_3} 
\nabla_{\beta_2}A_{\beta_3}$, since it is already included in ${\cal L}_2$.
For the sixth-order interaction, we run out of the indices to 
make the product $A^{\alpha_1}A_{\beta_1}$.
Instead, we consider the following Lagrangian density
\be
{\cal L}_6^{\rm N}
=\tilde{f}_{6}
 \hat{\delta}_{\alpha_1 \alpha_2 \alpha_3 \alpha_4}^{\beta_1 \beta_2\beta_3\beta_4}
\nabla_{\beta_1} A_{\beta_2} \nabla^{\alpha_1}A^{\alpha_2}
\nabla_{\beta_3} A^{\alpha_3} \nabla_{\beta_4} A^{\alpha_4}\,,
\label{L6N}
\ee
with $\tilde{f}_{6}(X)$. This is of the same form as Eq.~(\ref{L6V}), but the difference from 
generalized Proca theories is that the relative coefficient to the non-minimal coupling term 
$L^{\mu \nu \alpha \beta} \nabla_{\mu}A_{\nu} \nabla_{\alpha}A_{\beta}$ 
is detuned in beyond-generalized Proca theories, which generates derivatives higher than 
second order.
Then, the new Lagrangian densities in our set up are given by 
\be
{\cal L}^{\rm N}=
{\cal L}_4^{\rm N}+{\cal L}_5^{\rm N}+
\tilde{{\cal L}_5^{\rm N}}+
{\cal L}_6^{\rm N}\,.
\label{LN}
\ee
To study the effect of derivative interactions 
in beyond-generalized Proca theories, 
we consider the following action 
\be
S=\int d^4x \sqrt{-g} \left( \sum_{i=2}^{6} {\cal L}_i
+{\cal L}^{\rm N}+{\cal L}_M \right)\,, 
\label{action}
\ee
where ${\cal L}_M$ is the matter Lagrangian density.

In the following, we would like to analyze the possible number of 
propagating DOF in beyond-generalized Proca theories explained above.
The worry is that the new terms (\ref{LN}) might induce the propagation 
of a ghostly DOF associated with the Ostrogradski instability.
For this purpose, we shall focus on the study for both the background 
(Sec.~\ref{Hessiansec}) and the linear perturbation (Sec.~\ref{FLRWsec1})  
on top of the isotropic FLRW background.

Note that this first analysis does not necessarily guarantee the absence 
of ghostly DOF on more general backgrounds. 
For a complete proof of the absence of extra DOF, the full $3+1$ 
ADM Hamiltonian analysis is needed without fixing the gauge. 

\section{Background equations of motion and the Hamiltonian}
\label{Hessiansec}
\subsection{Background and perturbed quantities}

To derive the background and perturbation equations of motion 
on the isotropic cosmological background, 
we consider the general perturbed metric 
in the form \cite{Bardeen}
\ba
\hspace{-0.9cm}
& &
ds^{2}=-(1+2\alpha)\,dt^{2}+2\left( \chi_{|i}
+V_i \right)dt\,dx^{i} \nonumber\\
\hspace{-0.9cm}
&&+a^{2}(t) \left[ (1+2\psi) \delta_{ij}
+2E_{|ij}+2F_{i|j}
+h_{ij} \right] dx^i dx^j, 
\label{permet}
\ea
where $\alpha, \chi, \psi, E$ are scalar metric perturbations,
$V_i, F_i$ are vector perturbations, and $h_{ij}$ is the tensor perturbation. 
The index ``${}_{|}$'' represents the covariant 
derivative with respect to the three-dimensional spatial metric.
Expanding the action (\ref{action}) up to first order in scalar perturbations, 
we can obtain the background equations of motion on the flat FLRW 
background described by the line element 
$ds^2=-dt^2+a^2(t) \delta_{ij}dx^i dx^j$. 
The linear perturbation equations also follow from the action (\ref{action})
expanded up to second order in scalar, vector, and tensor perturbations.
Before doing so, we first remove redundant gauge DOFs.

Under a scalar gauge transformation 
$t \to t+\delta t$ and $x^i \to x^i+ \delta^{ij} \delta x_{|j}$, 
the scalar perturbations $\psi$ and $E$ transform, respectively, 
as $\psi \to \psi -H \delta t$ and $E \to E-\delta x$  \cite{Bassett}, where 
$H=\dot{a}/a$ is the Hubble expansion rate and a dot represents 
a derivative with respect to $t$.
Under a vector gauge transformation $x^i \to x^i+\delta x^i$, 
the vector perturbation $F_i$ transforms as 
$F_i \to F_i- \delta x_i$. 
If we choose the flat gauge 
\be
\psi=0\,,\qquad E=0,\qquad F_i=0\,,
\label{gauge}
\ee
then the time slicing $\delta t$, the spatial threading 
$\delta x$, and the infinitesimal vector $\delta x_i$ are 
unambiguously fixed. 

In what follows, we shall derive the equations of motion 
for the background and  cosmological 
perturbations under the gauge choice (\ref{gauge}). 
By fixing the gauge in this way, we already removed the extra gauge 
DOFs from the beginning.
We have also expanded the action (\ref{action}) up to second order 
in perturbations without fixing the gauge from the beginning and 
have derived the equations of motion from the general gauge-invariant Lagrangian.
Choosing the flat gauge (\ref{gauge}) in the equations of motion 
at the end, we confirmed that the resulting dynamical equations for tensor, vector, 
and scalar perturbations are equivalent to those derived by 
fixing the gauge from the beginning in the Lagrangian.

The vector perturbation satisfies the transverse condition 
$\partial^i V_i=0$, where $\partial^i$ represents the 
spatial derivative.
The tensor perturbation $h_{ij}$ obeys the transverse and
 traceless conditions $\partial^i h_{ij}=0$ and ${h_i}^i=0$. 
We express the temporal and spatial components of 
the vector field 
$A^{\mu}$, as 
\be
A^{0}=\phi(t)+\delta\phi\,,\qquad
A^{i}=\frac{1}{a^2(t)} \delta^{ij} \left(
\partial_{j}\chi_{V}+E_j \right)\,,
\ee
where $\phi(t)$ is the background value of the temporal 
vector component,  $\delta \phi$ and $\chi_{V}$ are the scalar 
perturbations, and $E_j$ is the intrinsic vector perturbation 
obeying the transverse condition $\partial^jE_j=0$.

For the matter sector,
we take into account a perfect fluid described by 
the Schutz-Sorkin action \cite{Sorkin}:
\be
S_{M}=-\int d^{4}x \left[ \sqrt{-g}\,\rho_M(n)
+J^{\mu} \left( \partial_{\mu}\ell+
\sum_{i=1}^{2} {\cal A}_i \partial_{\mu} {\cal B}_i
 \right) \right].
\label{Spf}
\ee
The energy density $\rho_M$ depends on the fluid number 
density $n=\sqrt{J_{\mu}J^{\mu}/g}$, where the temporal 
and spatial components of $J^{\mu}$ can be decomposed, 
respectively, as 
\be
J^{0}= \mathcal{N}_{0}+\delta J\,,\qquad
J^{i} =\frac{1}{a^2}\,\delta^{ik}
\left( \partial_{k}\delta j+W_k \right)\,,
\ee
where ${\cal N}_0$ is a constant associated with 
the total background particle number (related with 
the background number density $n_0$ as 
${\cal N}_0=n_0a^3$), $\delta J$ and $\delta j$ 
are the scalar perturbations, and $W_k$ is 
the vector perturbation satisfying $\partial^k W_k=0$.

The scalar quantity $\ell$ can be decomposed as 
$\ell=\ell_0-\rho_{M,n}v$, where the background 
value $\ell_0$ obeys the relation 
$\partial_0 \ell_0=-\rho_{M,n} \equiv 
-\partial \rho_M/\partial n$ and $v$ is the 
perturbation associated with the velocity potential.
Then, we can write $\ell$ in the form 
$\ell=-\int^t \rho_{M,n} (\tilde{t})\,d\tilde{t}
-\rho_{M,n}v$.

The terms ${\cal A}_i$ and ${\cal B}_i$ in Eq.~(\ref{Spf}) 
correspond to vector perturbations obeying 
the transverse conditions. It is sufficient to consider the 
$x,y$ components of ${\cal A}_i$ whose perturbations
depend on $t$ and $z$ alone, i.e., 
${\cal A}_1=\delta {\cal A}_1(t,z)$ and 
${\cal A}_2=\delta {\cal A}_2(t,z)$. 
One can extract the required property of the vector mode 
by choosing ${\cal B}_1=x+\delta {\cal B}_1(t,z)$ and 
${\cal B}_2=y+\delta {\cal B}_2(t,z)$.
Varying the matter action (\ref{Spf}) with respect to 
$J^{\mu}$, it follows that 
\be
J_{\mu}=\frac{n\sqrt{-g}}{\rho_{M,n}} 
\left( \partial_{\mu}\ell+
\sum_{i=1}^{2} {\cal A}_i \partial_{\mu} {\cal B}_i
 \right)\,,
\ee
which is related with the fluid four-velocity $u_{\mu}$, 
as $u_{\mu}=J_{\mu}/(n\sqrt{-g})$.
The spatial part of $u_{\mu}$ can be expressed as
\be
u_i=-\partial_{i}v+v_i\,,
\ee
where $v_i$ is the transverse vector perturbation 
associated with $\delta {\cal A}_i$, as
$\delta {\cal A}_i=\rho_{M,n}v_i$.

At the background level, the fluid action (\ref{Spf}) reads
\be
S_M^{(0)}=\int d^4 x \sqrt{-g}\, P_M(n_0)\,, \quad
P_M(n_0)=n_0\rho_{M,n}-\rho_M\,,
\label{PM}
\ee
where $P_M$ corresponds to the fluid pressure. 
As far as the scalar perturbation is concerned, the perfect fluid 
can be also described by the k-essence action \cite{kes}
\be
S_M=\int d^4 x \sqrt{-g}\,P_M(Z)\,, 
\quad Z =-\frac12 g^{\mu \nu} \partial_{\mu} \sigma
\partial_{\nu} \sigma\,,
\label{SMkes}
\ee
where the pressure $P_M$ depends on the kinetic term of 
a scalar field $\sigma$ (see also Refs.~\cite{kesori}).
At the background level the matter energy density
is given by $\rho_M=2ZP_{M,Z}-P_M$, so 
there is the correspondence 
$n_0\rho_{M,n} \to 2ZP_{M,Z}=\rho_M+P_M$.
{}From the k-essence action (\ref{SMkes}) we obtain the density 
perturbation $\delta \rho_M$, the pressure perturbation $\delta P_M$, and 
the velocity potential $v$, respectively, as 
\ba
\delta \rho_M&=&\left( P_{M,Z}+2ZP_{M,ZZ} \right)\delta Z\,, \nonumber\\
 \delta P_M&=&P_{M,Z} \delta Z\,, \nonumber\\
v&=&\frac{\delta \sigma}{\dot{\sigma}}\,,
\label{corres}
\ea
where $\delta Z$ corresponds to
\be
\delta Z=\dot{\sigma} \delta \dot{\sigma}
-\dot{\sigma}^2\alpha\,.
\label{deltaZ}
\ee
As far as the tensor and scalar perturbations are concerned, we can employ
either the Schutz-Sorkin action or the k-essence action, 
but for the computation of vector perturbations we need to resort 
to the Schutz-Sorkin action.

We shall expand the action (\ref{action}) 
together with the Schutz-Sorkin action (\ref{Spf}) up to second-order 
in perturbations on the flat FLRW background to discuss 
the propagating DOF.
In doing so, we perform the following field redefinitions:
\ba
Z_i&=&E_i+\phi(t)V_i\,,\nonumber\\
\psi&=&\chi_V+\phi(t)\chi\,,\nonumber\\
\delta \rho_M&=&\frac{\rho_{M,n}}{a^3} \delta J\,,
\label{Zidef}
\ea
where $Z_i$ and $\psi$ correspond to the vector and scalar 
parts of $A_i$ respectively, and $\delta \rho_M$ is the matter 
density perturbation. The vector field $Z_i$ obeys the 
transverse condition $\partial^iZ_i=0$, so there are 
two independent components.
At first order, the perturbation $\delta n$ of the fluid 
number density is equivalent to $\delta \rho_M/\rho_{M,n}$.

\subsection{Background equations}
\label{backeqsec}

Expanding the action (\ref{action}) up to first order 
in scalar perturbations, the resulting first-order action is given by 
\be
S^{(1)}=a^3 \left( {\cal C}_1 \alpha+
{\cal C}_2 \delta \phi+{\cal C}_3 v \right)\,,
\label{S1}
\ee
where we introduced the following short-cuts for convenience 
\ba
& &
\hspace{-0.4cm}
{\cal C}_1 
= G_2+G_{2,X}\phi^2+3G_{3,X}H \phi^3
+6(G_4+G_{4,XX}\phi^4)H^2  \nonumber \\
& &
\hspace{-0.4cm}
~~~-\left( G_{5,X}+G_{5,XX} \phi^2 \right)H^3 \phi^3 
-\rho_M
 \nonumber \\
& &
\hspace{-0.4cm}
~~~+6\left[ 3f_4+f_{4,X}\phi^2
+H\phi \left( 3f_5+f_{5,X} \phi^2 \right) \right]
H^2 \phi^4 \,,\label{beeq1} \\
& &
\hspace{-0.4cm}
{\cal C}_2 
= \phi \{ G_{2,X}+3G_{3,X}H\phi +6(G_{4,X}
+G_{4,XX}\phi^2)H^2
\nonumber \\
& &
\hspace{-0.4cm}
~~~-(3G_{5,X}+G_{5,XX}\phi^2) 
H^3\phi \nonumber\\
\hspace{-0.4cm}
&& + 6[4f_4+f_{4,X}\phi^2
+(5f_5+f_{5,X}\phi^2)H\phi]H^2 \phi^2 \}\,, \label{beeq2}\\
& &
\hspace{-0.4cm}
{\cal C}_3 
=-\frac{{\cal N}_0}{a^3}
\left( \dot{\rho}_{M,n}+3H \frac{{\cal N}_0}{a^3} \rho_{M,nn} 
\right)\,,\label{beeq3}
\ea
where $H=\dot{a}/a$ is the Hubble expansion rate.
Variations of the action (\ref{S1}) with respect to 
$\alpha, \delta \phi,v$ give rise to the 
background equations 
\be
{\cal C}_i=0\qquad (i=1,2,3),
\label{Cieq}
\ee
respectively. On using the properties ${\cal N}_0=n_0a^3$ and 
$n_0 \rho_{M,n}=\rho_M+P_M$, the third equation 
(${\cal C}_3=0$) corresponds to the matter continuity equation 
\be
\dot{\rho}_M+3H\left( \rho_M+P_M \right)=0\,.
\label{continuity}
\ee
In the k-essence description of the perfect fluid, the third 
term on the r.h.s. of Eq.~(\ref{S1}) is replaced by 
$a^3 P_{M,Z} \dot{\sigma} \dot{\delta \sigma}$.
Variation with respect to $\delta \sigma$ leads to 
the matter equation of motion 
$\frac{d}{dt}(a^3P_{M,Z}\dot{\sigma})=0$, i.e., 
\be
\left( P_{M,Z}+\dot{\sigma}^2 P_{M,ZZ} \right)\ddot{\sigma}
+3H P_{M,Z} \dot{\sigma}=0\,.
\label{continuity2}
\ee
Using the correspondence $\rho_M=2ZP_{M,Z}-P_M$, 
the continuity Eq.~(\ref{continuity}) follows from Eq.~(\ref{continuity2}). 

The terms containing $f_4$ and $f_5$ in Eqs.~(\ref{beeq1}) and 
(\ref{beeq2}) correspond to the new terms arising from 
the Lagrangians (\ref{L4N}) and (\ref{L5N}). 
They originate from the longitudinal component of the vector field, 
so it is expected that the equations of motion can be written 
in terms of the quantities similar to those appearing 
in GLPV theories \cite{GLPV}. To see the correspondence 
with GLPV theories, we introduce the following quantities
\ba
& &A_2=G_2\,,\qquad A_3=(2X)^{3/2} E_{3,X}\,, \nonumber\\
&& A_4=-G_4+2XG_{4,X}+4X^2f_4\,,\nonumber \\
& & A_5=-\sqrt{2}X^{3/2} \left( \frac13 G_{5,X}
-4Xf_5 \right)\,, \nonumber \\
 &&B_4=G_4\,, \qquad 
B_5=(2X)^{1/2}E_5\,, 
\label{ABre}
\ea
where $E_3(X)$ and $E_5(X)$ are auxiliary 
functions \cite{Gleyzes13} satisfying 
\be
G_3=E_3+2XE_{3,X}\,,\qquad 
G_{5,X}=\frac{E_5}{2X}+E_{5,X}\,.
\ee
Then, the two background equations ${\cal C}_1=0$ 
and ${\cal C}_2=0$ can be written in compacts forms:
\ba
& &
\hspace{-1.1cm}
A_2-6H^2 A_4-12H^3 A_5=\rho_M\,,
\label{back1}\\
& &
\hspace{-1.1cm}
\phi \left( A_{2,X}+3HA_{3,X}+6H^2 A_{4,X}
+6H^3 A_{5,X} \right)=0\,.
\label{back2}
\ea
Taking the time derivative of Eq.~(\ref{back1}) and 
using Eq.~(\ref{continuity}), it follows that 
\be
\dot{A}_3+4\dot{H}A_4+4H \dot{A}_4
+12H \dot{H}A_5+6H^2 \dot{A}_5=
\rho_M+P_M\,.
\label{back3}
\ee
The background Eqs.~(\ref{back1}) and (\ref{back3}) 
are of the same forms as those in GLPV theories 
(see Eqs.~(2.15) and (2.16) of Ref.~\cite{Kase14}) 
with the particular relation (\ref{back2}).
In GLPV theories the constraint (\ref{back2}) is absent, 
but in beyond-generalized Proca theories the relation (\ref{back2})
gives the constraint on the background 
trajectory with $\phi$ always related to 
$H$ \cite{DeFelice16} (e.g., analogous to the tracker 
solution \cite{DeFelice10} found for scalar Galileons).

{}From Eq.~(\ref{ABre}) there are two particular relations
\ba
A_4+B_4-2XB_{4,X}&=&4X^2f_4\,,\nonumber\\
A_5+\frac13 XB_{5,X}&=&(2X)^{5/2}f_5\,. 
\label{ABrelation}
\ea
In generalized Proca theories the Lagrangians ${\cal L}_4^{\rm N}$ 
and ${\cal L}_5^{\rm N}$ are absent, so that 
$f_4=0$ and $f_5=0$. In this case, 
the functions $B_4$ and $B_5$ are related with 
$A_4$ and $A_5$ according to the relations 
$A_4+B_4-2XB_{4,X}=0$ and $A_5+XB_{5,X}/3=0$.
In beyond-generalized Proca theories the functions $f_4$ 
and $f_5$ are non-zero, so there are two more free 
functions $B_4$ and $B_5$ than those in generalized 
Proca theories.
This situation is analogous to the extension 
of Horndeski theories to GLPV theories \cite{GLPV}.
We recall that the Lagrangians ${\cal L}_6$, 
$\tilde{{\cal L}}_5^{\rm N}$, and ${\cal L}_6^{\rm N}$, 
which correspond to the intrinsic 
vector mode, do not contribute to the background 
equations of motion.

Since the background Eqs.~(\ref{back1})-(\ref{back3}) 
do not contain the functions $B_{4}$ and $B_{5}$,
beyond-generalized Proca theories cannot be distinguished 
from generalized Proca theories at the background level 
(as it happens in the GLPV extension of Horndeski 
theories \cite{Kase14,KaseIJMPD}).
However, as we will discuss in Sec.~\ref{FLRWsec1}, 
this situation is different at the level of 
cosmological perturbations.

\subsection{Hamiltonian}

The discussion in Sec.~\ref{backeqsec} shows that, at the 
background level, beyond-generalized Proca theories do not 
give rise to additional ghostly DOF to that in generalized Proca theories.
It is also possible to see the absence of the Ostrogradski ghost
by computing the Hamiltonian of the system. 
In doing so, we consider the line element 
\be
ds^2=-N^2(t)dt^2+a^2(t) \delta_{ij}dx^idx^j\,,
\ee
which contains the lapse function $N(t)$. 
For the vector field given by $A^{\mu}=(\phi(t)/N(t),0,0,0)$,  
the action (\ref{action}) reduces to $S=\int d^4x\,L$, with
\ba
\hspace{-1.1cm}
& &L= Na^3 G_2-a^3 G_{3,X}\phi^2 \dot{\phi}
-\frac{6a\dot{a}^2G_4}{N}
+\frac{6a\dot{a}^2G_{4,X}\phi^2}{N} \nonumber\\
\hspace{-1.1cm}
&&~~~-\frac{G_{5,X}\dot{a}^3\phi^3}{N^2}
+\frac{6a \dot{a}^2f_4\phi^4}{N}
+\frac{6\dot{a}^3f_5 \phi^5}{N^2}+Na^3P_M,
\label{lag}
\ea
where we have carried out the integration by parts. 
Since the Lagrangian (\ref{lag}) does not contain the time 
derivative of $N$, there exists a Hamiltonian constraint. 
In fact, the variation of $L$ with respect to $N$ leads to 
\be
\frac{\partial L}{\partial N}=
-\frac{{\cal H}}{N}=0\,,
\label{con}
\ee
where ${\cal H}=\Pi^{\mu}\dot {\cal O}_\mu-L$ 
is the Hamiltonian with 
$\Pi^{\mu}=\partial L/\partial\dot {\cal O}_\mu$ 
and ${\cal O}_{\mu}=(N(t),\phi(t),a(t))$.
The explicit form of ${\cal H}$ is given by 
\ba
{\cal H}&=&-Na^3 \Big( G_2+6H^2 G_4-6G_{4,X}H^2 \phi^2
+2G_{5,X}H^3 \phi^3 \nonumber\\
&&-6f_4 H^2 \phi^4-12f_5H^3 \phi^5 
-\rho_M \Big)\,,
\label{Hami}
\ea
which does not contain any time derivatives of $\phi$.
Equation (\ref{con}) shows that ${\cal H}=0$ exactly. 
Hence there is no Ostrogradski instability associated 
with the Hamiltonian unbounded from below. 
Existence of the constraint (\ref{con}) removes 
the would-be ghostly DOF associated with the 
time derivatives of $\phi$.

The Hamiltonian constraint ${\cal H}=0$ follows from the background 
Eqs.~(\ref{Cieq}). In fact, after eliminating the term $G_{2,X}$ from the 
two equations ${\cal C}_{1}=0$, ${\cal C}_2=0$ and setting $N=1$, 
we obtain the constraint equation ${\cal H}=0$.
Moreover, varying the Lagrangian (\ref{lag}) with respect to $\phi$, 
the resulting equation of motion is equivalent to ${\cal C}_2=0$.

What we have shown in this section is by no means a full proof of 
the absence of extra ghostly DOF on arbitrary backgrounds. 
A full ADM Hamiltonian analysis is needed for this purpose. 
Even though this proof is not the goal of the present work, we will 
consider linear perturbations on the FLRW background in Sec. IV 
and investigate the propagating DOF.

\section{Dynamics of linear perturbations}
\label{FLRWsec1}

In this section we expand the action (\ref{action}) up to second 
order in tensor, vector, and scalar perturbations to study 
the number of DOFs as well as no-ghost and stability conditions 
for linear cosmological perturbations.

\subsection{Tensor perturbations}

We begin with the derivation of the second-order action 
for tensor perturbations $h_{ij}$. 
We can express $h_{ij}$ in terms of two polarization modes 
$h_{+}$ and $h_{\times}$, as 
$h_{ij}=h_{+}e_{ij}^{+}+h_{\times} e_{ij}^{\times}$.
The unit bases $e_{ij}^{+}$ and $e_{ij}^{\times}$ 
satisfy the normalization conditions
$e_{ij}^{+}({\bm k}) e_{ij}^{+}(-{\bm k})^*=1$,
$e_{ij}^{\times}({\bm k}) e_{ij}^{\times}(-{\bm k})^*=1$,
and $e_{ij}^{+}({\bm k}) e_{ij}^{\times}(-{\bm k})^*=0$
in Fourier space, where ${\bm k}$ is the comoving wave number. 
Expanding the action (\ref{action}) up to quadratic order 
in tensor perturbations, the second-order action reads
\be
S_T^{(2)}=\sum_{\lambda={+},{\times}}\int dt\,d^3x\,
a^3\,\frac{q_T}{8}  \left[\dot{h}_\lambda^2
-\frac{c_T^2}{a^2}(\partial h_\lambda)^2\right]\,,
\label{ST}
\ee
where
\ba
q_T &=& 2G_4-2G_{4,X}\phi^{2}+G_{5,X}H\phi^{3}
-2f_4\phi^4-6f_5H\phi^5 \nonumber\\
&=& -2 \left( A_4+3HA_5 \right)\,,
\label{qT}\\
c_T^2 &=& \frac{2G_{4}+G_{5,X}\phi^2 \dot{\phi}}{q_T} 
=-\frac{2B_4+\dot{B}_{5}}
{2(A_4+3HA_5)}\,.
\label{cT}
\ea
In the second equalities of Eqs.~(\ref{qT}) and (\ref{cT}) we 
have used the quantities defined by Eq.~(\ref{ABre}). 
The Lagrangians ${\cal L}_4^{\rm N}$ and 
${\cal L}_5^{\rm N}$ lead to the modification of 
$q_T$, which on the other hand can be expressed in terms of $A_4$ and 
$A_5$ alone. The numerator of $c_T^2$ contains the terms 
$B_4$ and $\dot{B}_{5}$, so beyond-generalized Proca theories give rise to 
the tensor propagation speed different from that in 
generalized Proca theories.
The expressions of $q_T$ and $c_T^2$ are 
of the same forms as those in GLPV theories \cite{KaseIJMPD}. 
The action (\ref{ST}) does not contain the derivative terms 
higher than second order, so the dynamical DOF of the tensor 
mode remain two.

\subsection{Vector perturbations}
\label{Vecsec}

Let us proceed to the discussion of vector perturbations.
Due to the transverse conditions of the vector mode 
(e.g., $\partial^iZ_i=0$), we can choose the components of 
these fields as $Z_i=(Z_1(t,z), Z_2 (t,z), 0)$ without 
losing the generality.
The second-order matter action $(S_M^{(2)})_V$ of the vector 
mode is the same as that derived in Refs.~\cite{DeFelice16,Geff}. 
Varying the action $(S_M^{(2)})_V$ with respect to 
$W_i, \delta {\cal A}_i, \delta {\cal B}_i$, 
we obtain the following relations 
\ba
W_i &=& {\cal N}_0 \left( v_i-V_i \right)\,,\\
\delta {\cal A}_i &=& \rho_{M,n}v_i=C_i\,,
\ea
where $C_i$ are constants in time, and 
\be
v_i=V_i-a^2\dot{\delta {\cal B}}_i\,.
\ee

After integrating out the fields $W_i$ and $\delta {\cal A}_i$, 
the full second-order action derived by expanding Eq.~(\ref{action}) 
in vector perturbations reads
\ba
& &
\hspace{-0.8cm}
S_V^{(2)} = \int dtd^3x\sum_{i=1}^{2} \biggl[
\frac{aq_V}{2} \dot{Z}_i^2
-\frac{1}{2a}\alpha_1 (\partial Z_i)^2
-\frac{a}{2\phi^2}\alpha_2Z_i^2 \nonumber\\
& &\hspace{-0.3cm}
+\frac{\phi}{2a}\alpha_3
\partial V_i \partial Z_i+\frac{q_T}{4a} (\partial V_i)^2
+\frac12 a(\rho_M+P_M)v_i^2 \biggr],
\label{Lv4}
\ea
where
\ba
q_V
&=& G_{2,F}+2G_{2,Y}\phi^2-4g_5H \phi
+2G_6H^2 \nonumber\\
&&+2G_{6,X} H^2 \phi^2+4\tilde{f}_6H^2\phi^2\,,
\label{qVdef} \\
\alpha_1
&=& q_V+2[ G_6\dot{H}-G_{2,Y}\phi^2
-\tilde{f}_5H \phi^3 \nonumber\\
& &-(H\phi-\dot{\phi})(G_{6,X}H\phi-g_5+
2\tilde{f}_6H\phi )
]\,,\\
\alpha_2
&=& 4G_4\dot{H}-4G_{4,X} H \phi \dot{\phi}
+2G_{5,X}H^2 \phi^2 \dot{\phi} \nonumber\\
& &+\rho_M+P_M\,,\\
\alpha_3
&=& 2G_{4,X}-G_{5,X}H \phi+2f_4\phi^2
+6f_5 H \phi^3 \nonumber\\
&=&\frac{2}{\phi^2}
\left( A_4+B_4+3HA_5 \right)\,.
\label{alpha3def}
\ea
The structure of the action (\ref{Lv4}) is the same as that 
derived in generalized Proca theories \cite{DeFelice16,Geff} 
with the different coefficients $q_V,\alpha_1,\alpha_2,\alpha_3$. 
Hence the new Lagrangians (\ref{L4N})-(\ref{L6N}) do not 
give rise to any additional DOF associated with vector perturbations.

Varying the action (\ref{Lv4}) with respect to $V_{i}$ yields
\ba
\frac{q_T}{2}
\frac{k^{2}}{a^{2}}V_{i}=-(\rho_M+P_M)v_{i}
-\frac{\alpha_3 \phi}{2}\frac{k^{2}}{a^{2}}Z_{i}\,,\label{vecre}
\ea
and similarly with respect to $Z_i$ :
\ba
& &
\ddot{Z}_i+\left( H+\frac{\dot{q}_V}{q_V} \right)\dot{Z}_i
+\frac{1}{q_V} \left( \alpha_1 \frac{k^2}{a^2}
+\frac{\alpha_2}{\phi^2} \right)Z_i \nonumber\\
& &-\frac{\alpha_3 \phi}{2q_V} \frac{k^2}{a^2}V_i=0\,.
\label{Zieq}
\ea
In the small-scale limit ($k \to \infty$) we can neglect the matter 
contribution in Eq.~(\ref{vecre}), so we obtain the 
approximate relation 
$V_i \simeq -(\alpha_3 \phi/q_T)Z_i$.
Substituting this into Eq.~(\ref{Zieq}), the 
dynamical vector field $Z_i$ obeys
\be
\ddot{Z}_i+\left( H+\frac{\dot{q}_V}{q_V} \right)\dot{Z}_i
+c_V^2\frac{k^2}{a^2}Z_i \simeq 0\,,
\ee
where the vector propagation speed squared is given by
\ba
\hspace{-0.3cm}
c_V^2&=& 
\frac{\alpha_3^2 \phi^2}{2q_Tq_V}+\frac{\alpha_1}{q_V} 
\nonumber \\
\hspace{-0.3cm}
&=& 1+\frac{2(A_4+B_4+3HA_5)^2}{\phi^2 q_T q_V} \nonumber \\
\hspace{-0.3cm}
&&+\frac{2(G_6\dot{H}-G_{2,Y}\phi^2
-\tilde{f}_5H \phi^3)}{q_V}   \nonumber \\
\hspace{-0.3cm}
&&-\frac{2(H\phi-\dot{\phi})(G_{6,X}H\phi-g_5+
2\tilde{f}_6H\phi )}{q_V}.
\ea
To avoid the ghost and the Laplacian instability 
on small scales, 
we require the conditions $q_V>0$ and $c_V^2>0$. 
All the new Lagrangian densities (\ref{L4N})-(\ref{L6N}) 
affect $c_V$ through the changes of coefficients 
(\ref{qVdef})-(\ref{alpha3def}), while $q_V$ is only 
modified by the term ${\cal L}_6^{\rm N}$.
In spite of these modifications, the DOF of vector 
perturbations remain two as those in 
generalized Proca theories.

\subsection{Scalar perturbations}
\label{scalarsec}

For scalar perturbations, we first expand the Schutz-Sorkin
action (\ref{Spf}) up to second order by using the matter 
perturbation $\delta \rho_M$ defined in Eq.~(\ref{Zidef}). 
Varying this action with respect to $\delta j$, we obtain 
\be
\partial \delta j=-a^3 n_0 \left( \partial v+\partial \chi 
\right)\,.
\ee
On using this relation and the background equation of motion, 
the second-order matter action 
reduces to 
\ba
(S_M)_S^{(2)}=\int dtd^3x\,a^3 \Big[ -\frac{n_0 \rho_{M,n}}{2a^2}(\partial v)^{2}+n_0\rho_{M,n}
v \frac{\partial^2 \chi}{a^2} \nonumber\\
+\dot{v}\,\delta \rho_M 
-3Hc_M^2v\,\delta \rho_M -\frac{c_M^2}
{2n_0 \rho_{M,n}}(\delta \rho_M)^2-\alpha \delta \rho_M \Big], \;\;\;\;\;
\label{SM2}
\ea
where $c_M^2$ is the matter sound speed squared defined by 
\be
c_M^2 \equiv \frac{P_{M,n}}{\rho_{M,n}}
=\frac{n_0\rho_{M,nn}}{\rho_{M,n}}\,.
\label{cM}
\ee
Expansion of the full action (\ref{action}) up to second order 
in scalar perturbations gives
\ba
S_{S}^{(2)}  &=&  \int dt d^3 x\,a^{3}\,\Biggl\{
\left(w_{1}\alpha+\frac{w_{2} \delta\phi}{\phi} \right)\frac{\partial^{2}\chi}{a^2}-w_{3} 
\frac{(\partial \alpha)^{2}}{a^{2}} \nonumber\\
&&+w_{4}\alpha^{2}
-\frac{w_3}{4}\,\frac{(\partial\delta\phi)^2}{a^2 \phi^2}
+w_{5} \frac{(\delta\phi)^{2}}{\phi^2}
-\frac{w_{3}}{4\phi^{2}}\,\frac{(\partial\dot{\psi})^{2}}{a^{2}}
\nonumber \\
&&+\frac{w_{7}}{2}\,
\frac{(\partial\psi)^{2}}{a^{2}}-(3Hw_{1}-2w_{4})
\alpha \frac{\delta\phi}{\phi}  \nonumber \\
&&+\alpha\left[w_{3}\,\frac{\partial^{2}(\delta\phi)}
 {a^{2}\phi}+w_{3}\,\frac{\partial^{2}\dot{\psi}}{a^{2}\phi}
 -w_{6}\,\frac{\partial^{2}\psi}{a^{2}}\right]  \nonumber\\
&&-\left( w_8\psi-w_{3}\dot{\psi} \right)
 \frac{\partial^{2}(\delta\phi)}{2a^{2}\phi^{2}} \Biggr\} +(S_M)_S^{(2)}\,, 
\label{sscalar}
\ea
where we introduced the following variables for compactness 
\ba
w_{1} & = & -A_{3,X}\phi^2+4H(A_4-A_{4,X}\phi^2) \nonumber\\
&&+6H^2 (2A_5-A_{5,X}\phi^2)\,,
\label{w1}\\
w_{2} & = & w_1+2Hq_T\,,\label{w2}\\
w_{3} & = & -2{\phi}^{2}q_V\,,\label{w3} \\
w_{4} & = & 3H(w_2-Hq_T)+w_5\,,\label{w4} \\
w_{5} & = & \frac12 \phi^4 \Big( A_{2,XX}
+3HA_{3,XX}  \nonumber\\
&&+6H^2A_{4,XX}+6H^3 A_{5,XX} \Big)\,,
\label{w5} \\
w_{6} & = & -\frac{1}{2\phi} 
\left[ 4H(q_T-2B_4)-w_8 \right]\,,
\label{w6} \\
w_{7} & = & \frac{2(q_T-2B_4)}{\phi^2} \dot{H}
+\frac{w_8}{2\phi^3}\dot{\phi}\,,
\label{w7}\\
w_{8} & = & 2w_2+4H\phi^2 \left( 2B_{4,X}
-H B_{5,X} \right)\,.
\label{w8}
\ea
The coefficients $w_1, w_2, w_4, w_5$ only contain the functions $A_i$ and their derivatives, but there exist the functions 
$B_{4,5}$ and their derivatives in $w_6, w_7, w_8$.
Hence the difference from generalized Proca theories arises 
through the terms containing $w_6, w_7, w_8$. 
In particular we have the following relation 
\ba
w_8-(w_6\phi+w_2) &=& 
-4H (A_4+B_4-2XB_{4,X})  \nonumber \\
&&-4H^2 (3A_5+XB_{5,X}) \nonumber \\
&=&-4H\phi^4 \left( f_4+3H\phi f_5 \right)\,,
\label{wre}
\ea
where we have used Eq.~(\ref{ABrelation}). 
In generalized Proca theories studied 
in Refs.~\cite{DeFelice16,Geff} 
we have that $f_4=f_5=0$, 
so there is the specific relation $w_8=w_6\phi+w_2$.
In beyond-generalized Proca theories, $w_8$ is different from $w_6\phi+w_2$.

While the presence of the Lagrangians ${\cal L}_4^{\rm N}$ 
and ${\cal L}_5^{\rm N}$ manifests themselves 
through the modifications of the functions $B_{4,5}$,
the effect of ${\cal L}_6^{\rm N}$ arises through the modification of the term $w_3=-2\phi^2 q_V$. 
The existence of $\tilde{{\cal L}}_5^{\rm N}$ does not 
affect the second-order action of scalar perturbations.

The structure of the action (\ref{sscalar}) is the same 
as that in generalized Proca theories derived 
in Refs.~\cite{DeFelice16,Geff}, so the new Lagrangian densities
(\ref{L4N})-(\ref{L6N}) do not give rise to any 
additional DOF. 
As in the GLPV extension of Horndeski theories \cite{GLPV}, 
there are no derivatives higher than second order 
in the scalar action (\ref{sscalar}).
Since this second-order property also holds for tensor and 
vector perturbations, beyond-generalized Proca theories 
with the new terms (\ref{L4N})-(\ref{L6N}) are not 
prone to the Ostrogradski instability on the flat FLRW background.

As we will see in the following, beyond-generalized Proca theories can be distinguished 
from generalized Proca theories by different evolution of 
the scalar propagation speed $c_S$. This situation should be 
analogous to that in GLPV theories where the new Lagrangians
beyond the Horndeski domain lead to the mixing 
between $c_S$ and the matter sound 
speed $c_M$ \cite{Gergely,GLPV}.
In order to see such a mixing explicitly, it is convenient to 
employ the k-essence description (\ref{SMkes}) 
of the perfect fluid. On using the correspondence (\ref{corres}) 
and the field equation of motion (\ref{continuity2}), the 
second-order matter action (\ref{SM2}) is equivalent to 
\ba
& &
\hspace{-0.4cm}
(S_M)_S^{(2)}
=\int dt d^3xa^3 \biggl[
\frac12(P_{M,Z}+\dot{\sigma}^2 P_{M,ZZ}) 
( \dot{\delta \sigma}^2 
-2\dot{\sigma} \alpha \dot{\delta \sigma} ) \nonumber \\
&&\qquad \quad~
-\frac{1}{2a^2} P_{M,Z} \left[ (\partial \delta \sigma)^2
+2\dot{\sigma} \partial \chi \partial \delta \sigma \right] 
\nonumber \\
& &
\qquad \quad~
+\frac12 \dot{\sigma}^2 \left(P_{M,Z}+\dot{\sigma}^2 
P_{M,ZZ} \right) \alpha^2 \biggr]\,.
\label{SMkes2}
\ea
The last term of Eq.~(\ref{SMkes2}) gives rise to the 
contribution to the term $w_4 \alpha^2$ in Eq.~(\ref{sscalar}). 
One can confirm that direct expansion of the k-essence action (\ref{SMkes}) leads to the second-order action 
same as Eq.~(\ref{SMkes2}).

On using Eq.~(\ref{SMkes2}) and varying the full action (\ref{sscalar}) with respect to 
$\alpha, \chi, \delta \phi$ respectively, we obtain 
the perturbation equations of motion in Fourier space:
\ba
& &
\hspace{-0.8cm}
\left(P_{M,Z}+\dot{\sigma}^2P_{M,ZZ} \right) 
( \dot{\sigma} \dot{\delta \sigma}-\dot{\sigma}^2 
\alpha)
\nonumber\\
& &
\hspace{-0.8cm}
+\left( 3Hw_1-2w_4 \right)\frac{\delta \phi}{\phi} -2w_4 \alpha \nonumber\\
&&
\hspace{-0.8cm}+\frac{k^2}{a^2} \left[ \frac{w_3}{\phi}
\left( \dot{\psi}+\delta \phi+2\alpha \phi \right)
+w_1 \chi-w_6 \psi \right]=0\,,
\label{per1} \\
& &
\hspace{-0.8cm}
P_{M,Z} \dot{\sigma} \delta \sigma+
w_1 \alpha+\frac{w_2}{\phi} \delta \phi=0\,, \label{per2}\\
& &
\hspace{-0.8cm}
\left( 3Hw_1-2w_4 \right)\alpha-2w_5 \frac{\delta \phi}{\phi}   \nonumber\\
&&
\hspace{-0.8cm}
+\frac{k^2}{a^2} \left[ \frac{w_3}{2\phi} \left( \dot{\psi}+\delta \phi+2\alpha \phi \right)
+w_2 \chi-\frac{w_8}{2\phi} \psi
\right]=0\,. \label{per3} 
\ea
{}From Eq.~(\ref{corres}) the first contributions to 
Eqs.~(\ref{per1}) and (\ref{per2}) can be written as
$(P_{M,Z}+\dot{\sigma}^2P_{M,ZZ}) 
(\dot{\sigma} \dot{\delta \sigma}-\dot{\sigma}^2 
\alpha)=\delta \rho_M$ and $P_{M,Z} \dot{\sigma} \delta \sigma=(\rho_M+P_M)v$, respectively.
On using Eqs.~(\ref{per1})-(\ref{per3}), we can express 
the perturbations $\alpha, \chi, \delta \phi$ in terms of 
$\psi, \delta \sigma$ and their derivatives.
Substituting those relations into Eq.~(\ref{sscalar}), 
the second-order scalar action reduces to 
the following form
\ba
& &
\hspace{-0.5cm}
S_{S}^{(2)}=\int dt d^3x\,a^{3}\Big( \dot{\vec{\mathcal{X}}}^{t}{\bm K}
\dot{\vec{\mathcal{X}}}
+\frac{k^2}{a^2}\vec{\mathcal{X}}^{t}{\bm G}
\vec{\mathcal{X}} \nonumber\\
& &
\hspace{-0.5cm}
\qquad \qquad \qquad\qquad
 -\vec{\mathcal{X}}^{t}{\bm M}
\vec{\mathcal{X}}
-\vec{\mathcal{X}}^{t}{\bm B}
\dot{\vec{\mathcal{X}}}
\Big) \,,
\label{SSfinal}
\ea
where ${\bm K}$, ${\bm G}$, ${\bm M}$, ${\bm B}$
are $2 \times 2$ matrices (${\bm M}$ does not 
contain the $k^2$ term), and the vector field
$\vec{\mathcal{X}}$ is defined by
\be
\vec{\mathcal{X}}^{t}=\left( \psi, \delta \sigma \right) \,.
\ee
The form of the action (\ref{SSfinal}) explicitly shows that 
there are only two scalar DOF coming from 
the field $\psi$ and the matter field $\delta \sigma$.

In the small-scale limit ($k \to \infty$), the components of 
the matrices ${\bm K}$ and ${\bm G}$ are given by\footnote{If we use the Schutz-Sorkin action 
itself for the matter sector, the leading-order contributions to 
$K_{22}$ and $G_{22}$ are proportional to $1/k^2$. 
After transforming the Schutz-Sorkin action to the k-essence 
action, both $K_{22}$ and $G_{22}$ do not have the 
$k$-dependence as the components $K_{11}$ and $G_{11}$. }
\ba
K_{11}&=&Q_S+\xi_1^2 K_{22}, \nonumber\\
K_{22}&=&\frac12 \left( P_{M,Z}+\dot{\sigma}^2P_{M,ZZ} 
\right)\,,\nonumber\\
K_{12}&=&K_{21}=\xi_1 K_{22}\,,
\ea
and 
\ba
G_{11}&=&{\cal G}+\dot{\mu}+H\mu\,,\nonumber\\
G_{22}&=&\frac12 P_{M,Z}\,,\nonumber\\
G_{12}&=&G_{21}=\xi_2 G_{22}\,,
\ea
where we introduced the following quantities
\ba
\hspace{-0.5cm}
Q_S&=&\frac{H^2q_T(3w_1^2+4q_Tw_4)}
{(w_1-2w_2)^2\phi^2}\,,\nonumber\\
\hspace{-0.5cm}
\xi_1&=&-\frac{w_2\dot{\sigma}}{(w_1-2w_2)\phi}\,,
\nonumber\\
\hspace{-0.5cm}
\xi_2&=&-\frac{(w_8-w_6\phi)\dot{\sigma}}{(w_1-2w_2)\phi}\,,
\nonumber \\
\hspace{-0.5cm}
{\cal G}&=&\frac{w_1w_8(4w_2w_6\phi-w_1w_8)-4w_2^2w_6^2 \phi^2}
{4w_3(w_1-2w_2)^2\phi^2}-\frac{w_7}{2},
\nonumber \\
\hspace{-0.5cm}
\mu&=&\frac{2w_2w_6\phi-w_1w_8}{4(w_1-2w_2)\phi^2}\,.
\ea
Provided that the matrix ${\bm K}$ is positive definite, 
the scalar ghosts are absent. 
Under the no-ghost condition $K_{22}>0$ of the fluid, 
the positivity of ${\bm K}$ is ensured for $Q_S>0$.
Since the quantity $Q_S$ does not contain the term $q_V$, 
the no-ghost condition is not affected by the intrinsic vector mode. 
We also note that $Q_S$ is solely expressed in terms 
of the functions $A_{3,4,5}$ and their derivatives, so the no-ghost condition is similar to that in generalized Proca theories.

In the large $k$ limit, the dominant contributions to the 
second-order action (\ref{SSfinal}) are the first two terms, 
so the dispersion relation is given by 
\be
{\rm det} \left( c_S^2 {\bm K}-{\bm G} 
\right)=0\,,
\ee
where $c_S^2$ is the sound speed squared 
related with the frequency $\omega$, as 
$\omega^2=c_S^2k^2/a^2$. 
Then, $c_S^2$ is the solution to the equation 
\be
\left(c_S^2 K_{11}-G_{11} \right)
\left(c_S^2 K_{22}-G_{22} \right)
-\left(c_S^2 K_{12}-G_{12} \right)^2=0\,.
\label{csso}
\ee
In generalized Proca theories there is the relation 
$w_8=w_6\phi+w_2$ and hence $\xi_1=\xi_2$. 
Since in this case $G_{12}/K_{12}=G_{22}/K_{22}$, 
we obtain the two decoupled solutions to Eq.~(\ref{csso}):
\ba
\hspace{-0.5cm}
c_M^2 &=&\frac{G_{22}}{K_{22}}=
\frac{P_{M,Z}}{P_{M,Z}+\dot{\sigma}^2 P_{M,ZZ}}\,,
\label{cm2} \\
\hspace{-0.5cm}
c_{\rm P}^2 &=& \frac{1}{Q_S} \left[ 
G_{11}-(K_{11}-Q_S)\frac{G_{22}}{K_{22}} \right] \nonumber\\
\hspace{-0.5cm}
&=&\frac{1}{Q_S} \left[ {\cal G}+
\dot{\mu}+H\mu-\frac{w_2^2 (\rho_M+P_M)}
{2(w_1-2w_2)^2\phi^2}\right]\,.
\label{cp2}
\ea
where $c_M^2$ is the matter propagation speed squared 
equivalent to Eq.~(\ref{cM}).
Another sound speed squared $c_{\rm P}^2$ 
coincides with the one derived in Refs.~\cite{DeFelice16,Geff}.

In beyond-generalized Proca theories we have that 
$\xi_1 \neq \xi_2$, in which case there is a mixing 
between the two scalar propagation speeds. 
To quantify the deviation from generalized Proca theories, 
we introduce the following dimensionless quantities 
\be
\alpha_{\rm P} \equiv \frac{\xi_2}{\xi_1}-1
=\frac{w_8-(w_6\phi+w_2)}{w_2}\,,
\label{alphaP}
\ee
and
\ba
\beta_{\rm P} &\equiv& 2c_M^2 \left( \frac{K_{11}}{Q_S} -1 
\right)\alpha_{\rm P} \nonumber\\
&=&\frac{w_2(w_8-w_6\phi-w_2)(\rho_M+P_M)}
{(3w_1^2+4q_Tw_4)q_TH^2}\,.
\label{bep}
\ea
Expressing the terms $G_{22},G_{11},K_{11},K_{22},G_{12}$ 
in terms of $c_M^2$ etc by using Eqs.~(\ref{cm2}), (\ref{cp2}), (\ref{bep}) as well as the relations $K_{22}=K_{12}^2/(K_{11}-Q_S)$ and $G_{12}/K_{12}=(1+\alpha_{\rm P})
G_{22}/K_{22}$, the two solutions to Eq.~(\ref{csso}) 
are given by 
\be
c_S^2=
\frac12 \left[ c_M^2+c_{\rm P}^2-\beta_{\rm P}
\pm \sqrt{(c_M^2-c_{\rm P}^2+\beta_{\rm P})^2+
2c_M^2 \alpha_{\rm P} \beta_{\rm P}} \right].
\label{cS2}
\ee
For non-relativistic matter ($c_M^2=0$), the two 
solutions (\ref{cS2}) reduce to $c_S^2=0$ and 
$c_S^2=c_{\rm P}^2-\beta_{\rm P}$. 
The latter corresponds to the scalar sound speed 
squared associated with the field $\psi$, whose 
value is different from $c_{\rm P}^2$ 
by the factor $\beta_{\rm P}$.
Thus, the sound speed squared is a 
key quantity to distinguish between  
generalized Proca theories and 
beyond-generalized Proca theories.
This situation is analogous to the difference
between Horndeski and GLPV theories. 
Since the contribution of the intrinsic vector mode affects 
$c_S^2$ through the term $w_3$ 
in the quantity ${\cal G}$, the sound speed 
in beyond-generalized Proca theories
generally differs from that in GLPV theories.

\section{Conclusions}
\label{consec}

We proposed the new derivative interactions 
(\ref{L4N})-(\ref{L6N}) beyond the domain of 
second-order generalized Proca theories.
These Lagrangian densities are constructed in terms 
of the products of the anti-symmetric Levi-Civita 
tensor as well as the vector field $A^{\mu}$ and 
its first derivatives. By taking the scalar limit 
$A^{\mu} \to \nabla^{\mu}\pi$, the terms 
${\cal L}_4^{\rm N}$ and 
${\cal L}_5^{\rm N}$ recover  
derivative interactions of the scalar field $\pi$ 
appearing in GLPV theories. 
Since it is known that GLPV theories do not increase 
the number of propagating DOF relative to that in 
Horndeski theories \cite{Hami,counting}, our interest is to see 
what happens by extending generalized Proca 
theories to those containing the four new interactions
(\ref{L4N})-(\ref{L6N}).

We first derived the dynamical equations of motion on 
the flat FLRW background and the associated Hamiltonian 
of the system. Even in the presence of new interactions,  
there is no additional ghostly DOF to that appearing in 
second-order generalized Proca theories.
In fact the existence of a constraint leads to a vanishing 
Hamiltonian, which explicitly shows the absence of 
the Ostrogradski ghost.

As a second step, we considered linear cosmological perturbations 
on top of the flat FLRW background by taking into 
account a perfect fluid and studied 
the propagating DOF by expanding the action 
(\ref{action}) up to second order in perturbations.
We showed that the number of DOF is the same as that 
in generalized Proca theories: two tensor polarizations, two transverse 
vector modes, and two scalar modes (one longitudinal scalar and 
one matter fluid). Thus, beyond-generalized Proca theories are not 
prone to the appearance of additional DOF at the level of linear 
cosmological perturbations.

We also found that the four new interactions affect the vector propagation 
speed squared $c_V^2$, while the vector 
no-ghost condition is only modified by the term 
${\cal L}_6^{\rm N}$.
By introducing the quantities given by Eq.~(\ref{ABre}),
we obtained the two relations (\ref{ABrelation}) analogous to 
those appearing in the GLPV extension of Horndeski theories.
Since the functions $f_4$ and $f_5$ do not vanish in beyond-generalized Proca 
theories, this leads to the scalar sound speed squared $c_S^2$ 
away from the value $c_{\rm P}^2$ of generalized Proca theories 
with the difference weighed by $\beta_{\rm P}$.
Thus, the two theories can be distinguished from each 
other by the different evolution of scalar and vector 
sound speeds.

There are several issues we did not address in this Letter.
While we showed that the number of DOF in beyond-generalized Proca theories 
is the same as that in generalized Proca theories on the FLRW 
background, it remains to see whether the same conclusion 
also holds at the fully non-linear level on general curved 
backgrounds. In doing so, it will be convenient to express 
the action (\ref{action}) in terms of quantities appearing in 
the 3+1 ADM decomposition of space-time (along 
the line of Ref.~\cite{Gleyzes13}).
In fact, we showed that the quantities associated with the FLRW 
background and tensor perturbations in beyond-generalized Proca 
theories can be expressed in simple 
forms by using the variables (\ref{ABre}) similar to those 
appearing in the ADM formulation of GLPV theories, 
but the situation is more involved for vector and scalar 
perturbations. In our case, there should be 
new contributions to the ADM action of GLPV theories associated 
with the vector mode. Moreover, it will be of interest to study 
the cosmological viability of dark energy models in the framework 
of beyond-generalized Proca theories. These topics will be left for future works.

\section*{Acknowledgements}

LH thanks financial support from Dr.~Max R\"ossler, 
the Walter Haefner Foundation and the ETH Zurich
Foundation.  
RK is supported by the Grant-in-Aid for Research Activity
Start-up of the JSPS No.\,15H06635. 
ST is supported by the Grant-in-Aid for Scientific Research Fund of 
the JSPS Nos.~24540286, 16K05359, and MEXT KAKENHI Grant-in-Aid for 
Scientific Research on Innovative Areas ``Cosmic Acceleration'' 
(No.\,15H05890).

\end{document}